\documentclass[useAMS,usenatbib]{mn2e}
\usepackage{epsfig}
\usepackage{psfig} 
\title[Starburst radio galaxies]
{Starburst radio galaxies: general properties, evolutionary histories and triggering}
\author[Tadhunter et al.]{C. Tadhunter$^{1}$, J. Holt$^{7}$,  
R. Gonz\'alez Delgado$^{2}$, J. Rodr\'iguez Zaur\'in$^{3}$,   
\newauthor
M. Villar-Mart\'in$^{2}$,
R.Morganti$^{4}$, B. Emonts$^{5}$, C. Ramos Almeida$^{1}$, K. Inskip$^{6}$
\\
\\
$^{1}$Department of Physics and Astronomy, University of Sheffield,  Sheffield, S3 7RH, UK\\ 
$^{2}$Instituto de Astrof\'isica de Andaluc\'ia (CSIC), Apdo. 3004, 18080 Granada, Spain \\
$^{3}$DAMIR, Instituto de Estructura de la Materia, CSIC, Madrid, Spain \\
$^{4}$ASTRON, PO Box 2, 7990 AA Dwingeloo, The Netherlands \\
$^{5}$CSIRO Astronomy and Space Science, Australia Telescope National Facility, PO Box 76, Epping NSW, 1710,
Australia \\
$^{6}$Max-Planck-Institut f\"ur Astronomie, K\"onigstuhl 17, D-69117 Heidelberg,
Germany \\
$^{7}$Leiden Observatory, Leiden University, PO Box 9513, 2300 RA Leiden, The Netherlands}
\begin{document}

\maketitle
\label{firstpage}
\begin{abstract}
In this paper we discuss the results of a programme of spectral synthesis modelling of a sample
of starburst radio galaxies in the 
context of scenarios for the triggering of the activity and the evolution of the host galaxies. 
New optical spectra are also presented for a subset of the objects discussed. The starburst
radio galaxies -- comprising $\sim$15 -- 25\% of all powerful extragalactic radio sources -- frequently show disturbed morphologies at optical wavelengths, and unusual
radio structures, although their stellar masses are typical of radio galaxies as a class. In terms of the characteristic ages of their young stellar populations (YSP), the objects can be divided into two groups:
those with YSP ages $t_{ysp} \le 0.1$~Gyr, in which the radio source has been triggered
quasi-simultaneously with the main starburst episode, and those with older YSP in which the radio source has been triggered or re-triggered a significant period after the starburst episode. Most of former group are associated with large mid- to far-IR (MFIR) continuum and [OIII] emission line luminosities 
($L_{IR} > 10^{11}$~L$_{\odot}$, $L_{[OIII} > 10^{35}$~W), while most of the latter have lower
luminosities. Combining the information on the YSP with that on the optical morphologies of the host galaxies, we deduce that the majority of the starburst radio galaxies have been triggered in galaxy mergers in which at least one of the galaxies is gas rich. However, the triggering (or re-triggering) of the radio jets can occur immediately before, around, or a significant period after the final coalescence of the merging nuclei, reflecting the complex gas infall histories of the merger events.
Although $\sim$25\% of starburst radio galaxies are sufficiently bright at MFIR wavelengths to be
classified as ultra-luminous infrared galaxies (ULIRGs), we show that only the most massive ULIRGs are capable of evolving into radio galaxies. Finally, for a small subset of starburst radio galaxies in rich clusters of galaxies,  cooling flows associated with the hot X-ray haloes offer a viable alternative to mergers as a trigger for the radio jet activity. Overall, our results provide further evidence that powerful radio jet activity can be triggered via a variety of mechanisms, including different evolutionary stages of major galaxy mergers; clearly radio-loud AGN activity is not solely associated with a particular stage of a unique type of gas accretion event.

\end{abstract}
\begin{keywords}
galaxies:active -- galaxies:jets -- galaxies:starburst -- galaxies:interactions 
\end{keywords}

\section{Introduction}

Active Galactic Nuclei (AGN) are increasingly recognised as a key element in the overall galaxy evolution process \citep[e.g.][]{cattaneo09}, yet the nature of the link between AGN and galaxy evolution is not understood in depth. 
The link can take two main forms: the triggering of AGN as gas is driven into the nuclear regions of galaxies as they evolve; and the direct impact of the AGN-driven outflows on the ISM in the galaxy bulges, haloes and any surrounding groups and clusters. The latter AGN feedback effect has been invoked to explain the correlations between black hole mass and host galaxy properties
\citep[e.g.][]{silk98,fabian99,dimatteo05,bower06}, as well as the high-end shape of the galaxy luminosity function \citep{benson03}. On the other hand, the triggering of AGN activity as a result of evolution-driven gas accretion can provide an explanation for the similarity between the redshift evolution of AGN activity and that of the global star formation rate of the field galaxy population \citep{dunlop90,madau96}. Clearly, if we are to incorporate the AGN feedback effect accurately into galaxy 
evolution models, it is important to understand how and when AGN are triggered as their host galaxies evolve. 



Deep ground-based imaging observations of powerful radio galaxies show morphological evidence  (tidal tails, fans, bridges and shells etc.) that a large fraction are triggered in galaxy interactions and mergers \citep{heckman86,smith89, ramos10}. 
A merger origin is also supported by the kinematics of the extended emission line regions in some sources \citep{tadhunter89,baum92}. 
However, the idea that the activity is always triggered in major galaxy mergers and interactions has been challenged by high resolution Hubble Space Telescope (HST) images which provide
evidence that the host galaxies of powerful AGN of both radio-loud and radio-quiet types are relatively quiescent
giant elliptical galaxies \citep{dunlop03}. 
At least part of this apparent discrepancy between the HST and ground-based results is likely to be a consequence of the fact that shallow HST observations are sensitive to high surface brightness
structures close to the nuclei, while the lower resolution ground-based studies are more sensitive to lower surface brightness structures on larger scales \citep[see discussion in][]{canalizo07,bennert08}.

While some radio galaxies show clear evidence for triggering in galaxy interactions and mergers, a significant subset have optical morphologies and emission line kinematics that do not support such a triggering mechanism. Many such cases are central cluster galaxies surrounded by massive haloes of hot gas \citep{tadhunter89,baum92}. Therefore the infall of warm/cool gas condensing from the
hot X-ray haloes in cooling flows has been been suggested as an alternative trigger for the AGN activity \citep[e.g.][]{bremer97}. In addition, based both on detailed X-ray studies of individual objects and optical studies of
large samples of low luminosity radio sources, the direct (Bondi) accretion of the {\it hot} ISM associated with the
X-ray haloes has been proposed to explain the fuelling of weak line radio galaxies (WLRG) in the local Universe 
\citep{best05,allen06,best06,hardcastle07,buttiglione10}.


Although the morphologies and emission line kinematics provide important information about the triggering events, this information is generally qualitative. For example, it is difficult to use such information to quantify when AGN are triggered in major galaxy mergers relative to the time of coalescence of the merging nuclei. Fortunately, another facet of the gas accretion events that trigger AGN activity
is that they are also likely to be associated with star formation in the host galaxies; therefore, by studying the young stellar populations in the predominantly early-type host galaxies of powerful radio sources, we can obtain key information about the nature of the triggering events.

Substantial recent progress has been made in quantifying the degree of recent star formation activity
in radio galaxies using both optical/UV observations of direct starlight 
\citep{tadhunter02,wills02,wills04,raimann05,baldi08} and far-IR observations of dust-reprocessed
starlight \citep{tadhunter07,dicken08,dicken09}. Concentrating on optical/UV spectroscopic studies 
of complete samples that
take full account of contamination of the continuum by AGN-related components (e.g. scattered or direct
AGN light, nebular continuum), as well as intrinsic reddening of the starlight, young stellar populations (YSP) are detected
spectroscopically in 15 -- 30\% of powerful 2Jy and 3CR radio galaxies at low and intermediate redshifts 
\citep{tadhunter02,wills02,wills04,holt07}; the major uncertainties in these numbers are 
due to objects in which the strength of the direct or scattered AGN component makes detection of the YSP difficult. Despite the possibility that a substantial proportion of YSP emission may be absorbed by dust
at optical/UV wavelengths, analysis of the mid- to far-IR (MFIR) continuum properties of powerful radio galaxies
derived from deep Spitzer observations yields a similar proportion of powerful radio galaxies with energetically
significant recent star formation activity ($\sim$15 -- 28\%: Tadhunter et al. 2007, Dicken
et al., 2008, 2009)\footnote{HST imaging observations at UV wavelengths show evidence for YSP in a higher proportion of radio galaxies with strong emission lines \citep{baldi08}. However, since the level of dilution by the light of the old stellar populations of the host galaxies is more than a factor $\times$10 less at
UV wavelengths than optical wavelengths, such observations are sensitive to relatively low levels of recent star formation activity that may be
insignificant in terms of their contribution to the masses and energetics of the host galaxies.}.
 

The identification of YSP in a significant fraction of nearby radio galaxies has opened the possibility of using spectral synthesis modelling of the stellar populations to investigate the AGN triggering events and the evolution of the host galaxies. Initial results provided a mixed picture. On the one hand, Tadhunter et al. (2005) and \citet{emonts06} found evidence for intermediate YSP ages in their small sample of four nearby radio galaxies,
pointing to a significant delay between the starbursts and the AGN activity, similar
to those reported for samples of
star forming galaxies and radio-quiet AGN in the local Universe \citep[e.g.][]{davies07,wild10}.
On the other hand, a detailed study by \citet{wills08} of two intermediate redshift radio galaxies demonstrated the presence
of much younger stellar populations in one object, as well as intermediate age stellar populations in both
objects. All the spectral synthesis modelling studies of radio galaxies have stressed the following common features.
\begin{itemize}
\item {\bf Masses.} The masses associated with the YSP detected spectroscopically in radio galaxies are significant ($1 \times 10^9 < M_{ysp} < 5 \times 10^{10}$~M$_{\odot}$), amounting to typically 1 -- 40\% 
of the total stellar masses of the host galaxies; if the starbursts and AGN are triggered in mergers, then the systems
with the more massive YSP must be associated with relatively major mergers, in which at least one of the merging galaxies is gas-rich.
\item {\bf Spatial distribution.} When substantial YSP are detected in the nuclear regions of the radio galaxies, in general they are also detected across the full extents of the galaxies over which it is possible to make spectroscopic observations with the high S/N required for spectral synthesis modellling \citep[e.g.][]{holt07}. In some
cases, the YSP are detected on scales of 10s of kpc. Again this is consistent with simulations of gas rich mergers that predict galaxy wide  star formation activity  \citep[e.g.][]{mihos96,barnes04}, although it does not preclude other triggering mechanisms.
\item {\bf Reddening.} Reddening of the optical/UV light of the YSP by dust is important, especially in the nuclear regions of the galaxies (typically $0.4 < E(B-V) < 1.0$); failure to take such reddening into account leads to inaccurate ages for the YSP and, in particular, to the over-estimation of the ages of the YSP.
\end{itemize}

In this paper we draw together and summarise the results of all the recent spectral synthesis studies
of radio galaxies that are based on high quality optical spectra, including the recent investigation of
12 low and intermediate redshift radio galaxies by \citet{holt07}. New spectral synthesis results are also presented for a subset of the sample. The properties of the YSP are then discussed in the context of the morphologies of the host galaxies, in order to investigate the nature, timescales, and order of events of the triggering of the activity of an important subset of 
the population of galaxies with powerful AGN. Throughout this paper we assume a cosmology with $H_0 = 72$~km s$^{-1}$, $\Omega_m = 0.27$, and $\Omega_{\Lambda} = 0.73$.

\section{Sample selection and spectral synthesis modelling}

Optical and MFIR observations provide evidence for recent star formation activity in
up to 30\% of all radio galaxies. However, in this paper we concentrate on the $\sim$15 -- 25\% of the radio galaxy population with clear, unambiguous evidence for young stellar populations
at optical wavelengths, which have been modelled using spectral synthesis techniques that
take full account of the reddening  and potential AGN contamination. These latter
criteria are important because of the known contamination by scattered or direct AGN light
and/or nebular continuum from the NLR in some objects, the strong evidence that the 
YSP in the nuclear regions are significantly reddened in many cases, and potential degeneracy issues when more than one stellar population is present \citep[see][for a full discussion of these issues]{tadhunter02, holt07, wills08}. Our sample comprises all 18 of the objects fulfilling these selection criteria from our own detailed 
studies of nearby 2Jy, 3CR and B2 radio galaxies 
\citep{tadhunter05,holt07,emonts06,wills08}\footnote{In all of these studies we used the instantaneous burst, solar metallicity spectral synthesis results of \citet{bruzual03} to model the spectra.},
along with 3C48, which has been modelled using a similar approach\footnote{Note that, although 
\citet{canalizo00} and \citet{stockton07} did not consider the effects of reddening
in their study of the YSP in this luminous quasar, the regions studied are well
separated from the nucleus and dominated by the light of young/intermediate age YSP. Therefore, degeneracy issues are likely to be less severe than in some near-nuclear regions where
old stellar populations make a larger contribution, and the ages derived using unreddened templates
are likely to be accurate.}. In addition, we include a further two nearby objects from the literature with good spectroscopic observations of individual star clusters (Centaurus A and
Fornax A); such clusters do no suffer from the degeneracy issues associated with multiple stellar populations and AGN contamination. For convienience we label all the objects in our sample
collectively as `starburst radio galaxies', with the caveat that in some cases the star formation histories
may have been more complex than a single starburst.

Although not in any sense complete, our sample includes all 7 (15\%) of the 2Jy sample of 46 southern radio galaxies with intermediate redshifts ($0.05 < z < 0.7$) described in Dicken et al.
(2008), and all 5 (25\%) of the low redshift ($z < 0.1$) sample of 19 3CRR FRII radio galaxies 
described in Dicken et al. (2010), that show strong evidence for recent star formation activity at optical wavelengths. Therefore it is representative of the most extreme star forming radio 
galaxies in the local Universe; in these objects the YSP typically contribute $>$20\% of the total light
at $\sim$4800\AA. We emphasise that lower levels of star formation activity may be present in a larger fraction of the local radio galaxy population, but not clearly detected at optical wavelengths because of dilution by the light of the old stellar populations of the host galaxies and/or AGN-related continuum components.
 
The general properties of the starburst radio galaxies are presented in Table 1, while the properties of their YSP, as derived from spectral synthesis modelling, are detailed in Table 2.
Note that, with the exception of the individual star clusters in Cen A and Fornax A, the results presented in Table 2 refer to the simplest models that provide
an adequate fit to the overall optical/near-UV SEDs and detailed absorption features \citep[see][for a detailed description of the modelling approach]{holt07,wills08}. These comprise a reddened YSP combined with an unreddened old stellar population (OSP: $t_{osp} > 5$ Gyr). Because of the potential degeneracy issues, related to the relatively large contribution from OSP and reddening of the YSP, more detailed fits are not, in general, warranted. The exception
is the case of 3C459 discussed by \citep{wills08} in which the dominance of the young stellar component relative to the OSP allows two YSP components of different age (young: $t_{ysp} < 0.1$~Gyr; and intermediate: $0.3 < t_{ysp} < 0.9$~Gyr) to be fitted to the data, although the results in Table 2 for that object refer to the single reddeneded YSP plus OSP fit. Overall, the
results presented in Table 2 represent the luminosity-weighted ages of the YSP present in each system.   
 
\begin{table*}
\begin{tabular}{llrrrll}
\hline \\
Object &Redshift &P$_{5GHz}$ &L$_{[OIII]}$ &L$_{ir}$ &Optical morphology &Radio morphology \\
& &(W Hz$^{-1}$) &(W) &($L_{\odot}$) & \\
\hline \\
PKS0023-26 &0.322 &$1.1\times10^{27}$ &$1.5\times10^{35}$ &$1.5\times10^{11}$ &E, A, CE &CSS\\
NGC612 &0.030 &$7.9\times10^{24}$ &$5.6\times10^{32}$ &$1.1\times10^{11}$ &S0, D, S &FRI/FRII\\
3C48 &0.367 &$3.6\times10^{26}$ &$6.8\times10^{36}$ &$1.0\times10^{13}$ &Q, 2N, A, TT &CSS\\
Fornax A &0.006 &$5.6\times10^{24}$ &$<6.8\times10^{31}$ &$6.0\times10^{9}$ &E/S0, D, S &FRII\\
PKS0409-75 &0.693 &$8.7\times10^{27}$ &$1.3\times10^{35}$ &$5.3\times10^{11}$ &E, 2N &FRII \\
B2~0648+27 &0.041 &$1.9\times10^{23}$ &$2.7\times10^{34}$ &$2.6\times10^{11}$ &E, R, DE, TT &CSS\\
PKS0620-52 &0.0511 &$7.4\times10^{24}$ &$<2.5\times19^{32}$ &$2.6\times10^{10}$ &E &FRI \\
3C213.1 &0.194 &$7.0\times10^{25}$ &$1.3\times10^{34}$ &$<1.9\times10^{11}$ &E, DE  &FRI/FRII, DD\\
3C218 &0.055 &$9.5\times10^{25}$ &$2.9\times10^{33}$ &$2.1\times10^{10}$ &E, D &FRI, HSBI+DO\\
3C236 &0.101 &$4.1\times10^{25}$ &$3.8\times10^{33}$ &$4.1\times10^{10}$ &E, S, D, TT, DE &FRII, DD\\
3C285 &0.079 &$9.2\times10^{24}$ &$7.0\times10^{33}$ &$1.1\times10^{11}$ &E, TT, F, B &FRII\\
Centaurus A &3.4Mpc &$5.4\times10^{23}$ &$1.1\times10^{34}$ &$1.0\times10^{10}$ &E/S0, D, S, R &FRI, HSBI+DO\\
PKS1345+12 &0.122  &$1.0\times10^{26}$ &$2.0\times10^{35}$ &$1.9\times10^{12}$ &E, 2N, D, TT &GPS, HSBI+DO\\
3C293 &0.045 &$9.0\times10^{24}$ &$2.9\times10^{32}$ &$3.7\times10^{10}$ &E/S0, TT, D, B &FRI/FRII, HSBI+DO\\
3C305 &0.041 &$4.3\times10^{24}$ &$7.0\times10^{33}$ &$3.2\times10^{10}$ &E/S0, TT, D &CSS \\
3C321 &0.096 &$2.7\times10^{25}$ &$1.3\times10^{35}$ &$6.0\times10^{11}$ &E, 2N, TT, S, F &FRII, SSC\\
PKS1549-79 &0.152 &$2.5\times10^{26}$ &$1.0\times10^{35}$ &$2.2\times10^{12}$ &E, TT, D &CSS/CFS\\
PKS1932-46 &0.231 &$5.4\times10^{26}$ &$2.5\times10^{35}$ &$6.3\times10^{10}$ &E, S, A &FRII\\
3C433 &0.106 &$9.5\times10^{25}$ &$5.0\times10^{34}$ &$3.3\times10^{11}$ &E, A, CE, D &FRI/FRII, SSC\\
PKS2135-209 &0.636 &$2.5\times10^{27}$ &$1.4\times10^{36}$ &$1.3\times10^{12}$ &Q, F &CSS\\
3C459 &0.220 &$1.8\times10^{26}$ &$1.6\times10^{35}$ &$1.7\times10^{12}$ &E, F, DE &FRII, SSC\\
\hline \\
\end{tabular}
\caption{The general properties of the objects discussed in this paper. The integrated mid- to far-IR (MFIR) luminosities 
presented in column 3 have been estimated from published IRAS and Spitzer flux data using the formulae given in 
\citet{sanders96}.
Key for the optical morphology: E -- elliptical; S0 -- S0 galaxy; A -- amorphous halo; CE common -- envelope
with companion galaxies; D -- dust lane; 2N -- double nucleus; TT -- tidal tail(s); S -- shell(s);
DE -- distorted envelope; B -- bridge to companion galaxy; F -- fan(s); R -- ring structure. Key for
radio morphology: FRI -- Fanaroff Riley class I; FRII -- Fanaroff Riley class II; CSS -- compact steep spectrum
radio source; GPS -- giga-hertz peak radio source; CFS -- compact flat spectrum radio source; DD -- double-double
radio structure; HSBI+DO -- high surface brightness inner structure plus diffuse outer structure; SSC -- steep spectrum core structure. Note that the radio morphological classifications were
made by CT, based on published radio maps.} 
\end{table*}
\begin{table*}
\begin{tabular}{llllll}
\hline \\
Object &Region &YSP age &YSP Reddening &Comment &References \\
& &($Gyr$) &E(B-V) & & \\
\hline \\
PKS0023-26 &N &0.03 -- 0.05 &0.9 -- 0.7 & &1 \\
NGC612 &N &0.05 -- 0.1 &1.4 -- 1.2 & &1 \\
	&E: A -- F &0.04 -- 0.2 &1.0 -- 0.0 &Extended disk &1 \\
3C48 &E: 3C48~A &0.14 &0 &Secondary nucleus &2 \\
	&E: A, B, C, G &0.005 -- 0.114 &-- &Several regions, 4 slits &3 \\
Fornax~A &E &2.5 -- 3.5 &-- &Luminous star clusters &4 \\
	&E &2.0 &-- &Diffuse light, line index work, luminosity weighted &5 \\
	&N, E &0 -- 2 &0 -- 1.5 &Diffuse light, SED modelling &6 \\
PKS0409-75 &N &0.01 -- 0.04 &0.9 -- 0.5 &-- &1 \\
B2~0648+27 &N &0.2 -- 0.4 &0.2 -- 0.4 &-- &7 \\
	&E: NE &0.3 -- 0.6 &0.0 -- 0.1 & &7 \\
	&E: SW &0.3 -- 0.6 &0.0 -- 0.1 & &7 \\
PKS0620-52 &N &0.001 -- 0.9 &0.0 -- 1.5 &-- &6 \\
3C213.1 &N &0.4 -- 0.8 &0.1 -- 0.0 &-- &10 \\
3C218  &N &0.05 &0.4 &-- &1 \\
3C236 &N &0.04 -- 1.0 &1.6 -- 0.6 &-- &1 \\
	&E &0.01, 0.1 -- 1 &-- &Extended star formation complexes, photometric &8 \\
3C285 &N &0.1 -- 0.5 &0.2 -- 0.0 &-- &1 \\
Centaurus~A &E &0.35 &-- &Blue tidal stream &12 \\
	&E &2 -- 8 &-- &Globular cluster system &13 \\
PKS1345+12 &N &$\le$0.06 &0.8 -- 1.3 &-- &15 \\
&E, E1 &0.5 -- 1.5 &-- &-- &14 \\
3C293 &N &1.0 -- 2.5 &0.5 -- 0.2 &-- &14 \\
      &E, E1 &0.1 -- 0.5 &-- &-- &14 \\
      &E, E2 &1.0 -- 2.5 &-- &-- &14 \\
3C305 &N &0.4 -- 1.0 &0.8 -- 0.4 &-- &14 \\
      &E1 &0.5 -- 1.5 &-- &-- &14 \\
      &E2 &0.5 -- 1.0 &-- &-- &14 \\
3C321 &N, SE &0.1 -- 1.0 &0.3 -- 0.1 &-- &1 \\
	&N, NW &0.1 -- 1.4 &0.3 -- 0.1 &Radio source host &1 \\
	&E &0.4 -- 1.0 &0.1 -- 0.0 &Extended regions between nuclei and to SE &1 \\
PKS1549-79 &N &0.04 -- 0.1 &0.75 -- 0.0 &-- &9 \\
	&E &0.04 -- 0.08  &0.4 -- 0.0 &Extended envelope to S &6 \\
PKS1932-46 &E &$<$0.01 &-- &Extended star forming halo detected in emission lines &11 \\
3C433 &N, SW &0.03 -- 0.1 &0.7 -- 0.4 &Radio source host &1 \\
	&N, NE &0.03 -- 1.4 &0.9 -- 0.3 &Companion galaxy &1 \\
	&E, mid &0.05 -- 1.0 &0.7 -- 0.0 &Mid way between 3C433 \& companion &1 \\
PKS2135-209 &N &0.04 -- 0.6 &0.5 -- 0.2 &Quasar nucleus modelled with power-law &1 \\
PKS2314+03 &N &0.04 -- 0.07 &0.4 -- 0.0 &-- &10 \\
	&E, N &0.5 -- 0.7 &0.3 -- 0.0 &Fit includes a power-law &10 \\
	&E, E &0.5 -- 0.6 &0.2 -- 0.0 &Fit includes a power-law &10 \\
\hline \\
\end{tabular}
\caption{The properties of the young stellar populations in the radio galaxies
listed in Table 1. The second column indicates the aperture (N: nuclear; E: extended).
The references in column 6 are for the modelling of the young stellar
populations: 1. \citet{holt07}; 2. \citet{stockton07}; 3. \citet{canalizo00}; 4. \citet{Goudfrooij01};
5. \citet{kuntschner00}; 6. this paper; 7. \citet{emonts06}; 8. \citet{odea01}; 9. \citet{holt06};
10. \citet{wills08}; 11. \citet{villar05}; 12. \citet{peng02}; 13. \citet{peng04}; 14. \citet{tadhunter05};
15. \citet{zaurin09}.
} 
\end{table*}

\section{New spectroscopic observations}

In addition to spectroscopic observations already reported in the literature and described in the Appendix, new spectra for some of the starburst radio galaxies were obtained in various runs
on the ESO Very Large Telescope (VLT), 3.6m and New Technology Telescope (NTT) telescopes. Details of these new observations are given in Table 3. The reduction of these data followed the standard steps of bias subtraction, flat fielding, 
wavelength calibration, atmospheric extinction correction, flux calibration, and
registration of blue and red spectral images. Based on measurements of night sky lines,
the uncertainty in the wavelength calibration is estimated to be smaller than
0.5\AA\, for all datasets; the relative flux calibration uncertainty, based on 
the comparison of observations of several spectroscopic standard
stars, is better than $\pm$5\%.
Prior to the analysis of the data, the spectra were corrected for Galactic
extinction using values of E(B-V) derived from \citet{schlegel98} along with the \cite{seaton79} extinction law. Note that, in the cases of the VLT/FORS1 observations of PKS0023-26 and PKS1549-79, the data were taken in spectropolarimetric mode with the light passing through a half-wave plate as
well as an analysing prism. In the latter two cases, the `o' and `e' ray images of the long-slit spectra were spatially registered and co-added, and the flux calibration was performed using spectropolarimetric standard stars observed through the same instrument configuration as the target galaxies. 

For some of the objects discussed in this paper (Fornax A, PKS0652-20, PKS1549-79) we have carried out new spectral synthesis modelling in an attempt to improve our knowledge of the properties of the young stellar populations (YSP). In these cases we followed the procedures outlined in \citet{tadhunter05}, \citet{holt07}, and \citet{wills08}:
modelling the continuum spectrum using a combination of a reddened YSP ($0.005 < t_{ysp} < 5$~Gyr, $0 < E(B-V) < 2.0$) and an unreddened OSP ($t_{ysp}=12.5$~Gyr), taking full account of AGN continuum components (where appropriate). In all cases we used instantaneous burst, solar metallicity spectral synthesis templates from

\citet{bruzual03}.
\begin{table*}
\begin{tabular}{lllllll} 
\hline \\
Object &Date &Tel/Inst &Grism &Exposure &Slit &Seeing \\
& & & &Time (s) &Width (arcsec) &FWHM (arcsec) \\
\hline \\
PKS0023-26 &24/07/03 &VLT/FORS1 &GRIS-600R &5580 &1.3 &0.70 \\
Fornax A &7/12/02 &NTT/EMMI &Gr5 &900 &1.0 &1.1 \\
&'' &'' &Gr6 &900 &'' &'' \\
PKS0620-52 &16/02/07 &ESO3.6m/EFOSC &Gr7 &3600 &1.2 &1.7 \\
&'' &'' &Gr4 &1800 &'' &'' \\
PKS1549-79 &24/07/03 &VLT/FORS1 &GRIS-600B &5880 &1.3 &0.65 \\
\hline \\
\end{tabular}
\caption{New spectroscopic observations.} 
\end{table*}

\section{General patterns and results}

Detailed discussions of the individual starburst radio galaxies, including the new spectroscopic results, are presented in the Appendix. In this section we consider the general results and patterns obtained for the population of starburst radio galaxies as a whole.

\subsection{Radio morphologies}

It is striking that a large proportion of starburst radio galaxies show unusual radio morphologies that place them
outside the regular FRI/FRII morphological classification for extended radio sources: 7 (33\%) are compact steep spectrum (CSS) or Gigahertz peaked (GPS) sources whose radio structures are dominated by structures with a diameter $D < 15$kpc (PKS0023-26, 3C48, B2~0648+27, PKS1345+12, 3C305, PKS1549-79, PKS2135-209); 3 (14\%) show unusually prominent compact steep spectrum core components on a scale $D < 10$kpc, even if their radio emission is dominated by radio lobes and hotspots on a larger scale (3C321, 3C433, 3C459); 6 (29\%) show inner high surface brightness steep spectrum structures along with lower surface brightness outer haloes or double structures (3C213.1, 3C218, 3C236, Cen A, PKS1345+12, 3C293); and 15 (71\%) show one or more of these peculiarities. For comparison, the rate of detection of such features in the southern 2Jy sample of radio galaxies with redshifts in the range $0.05 < z < 0.7$ (see Dicken et al. 2008 for sample definition) is only 28\%. 

Although our small sample size makes it difficult to find clear trends in the detailed radio morphologies with the properties of the stellar populations, it is notable that all but one (3C305) of the 7 CSS/GPS sources in our sample have nuclear spectra that are consistent with relatively young ages for their YSP ($t_{ysp} < 0.1$~Gyr). 

\subsection{Optical morphologies}

At optical wavelengths most of the objects in our sample of starburst radio galaxies show morphological peculiarities compared with quiescent elliptical galaxies: 17 (80\%) show tidal tails, fans, or highly asymmetric/clumpy
outer envelopes at relatively high surface brightness levels; 14 (67\%) show dust features; 5 (23\%) have double nuclei or close companions within 15~kpc; and 20 (95\%) show one or more of these optical peculiarities. This rate of incidence is much higher than in the general population of massive elliptical galaxies observed with similar surface brightness sensitivity \citep[e.g.][]{malin83b}. However, at this relatively crude level of morphological classification, a similar rate of morphological disturbance has recently been found in the general population of powerful 2Jy radio galaxies at intermediate redshifts (including non-starburst objects: Ramos Almeida et al. 2010).

\subsection{YSP ages} 

\begin{figure*} 
\psfig{file=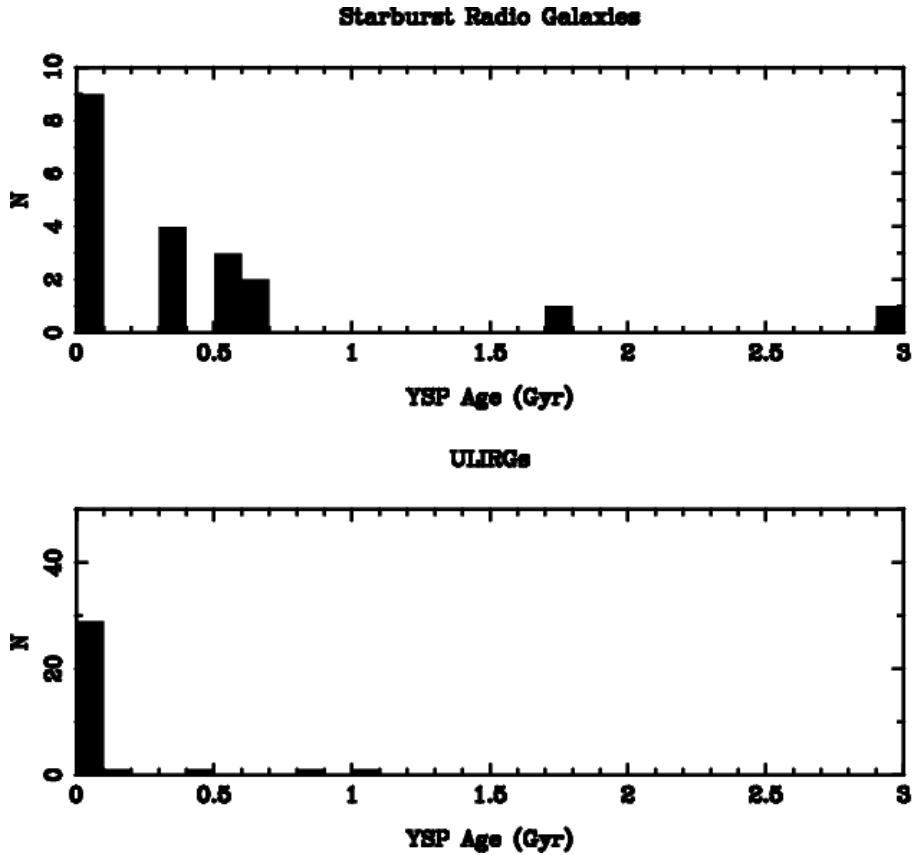, width=12cm}
\caption{The ages of the YSP in starburst radio galaxies (top), compared with
those of $z < 0.13$ ULIRGs
from \citet{zaurin09} (bottom).}
\end{figure*}

The ages of the YSP detected in radio galaxies can provide key information about the order-of-events and the triggering of the AGN/jet activity. Therefore it is interesting to examine the distribution of luminosity-weighted YSP ages determined from the two component fits to the optical spectra of the full sample of starburst radio galaxies
described in section 2. The top panel of Figure 1 shows the distribution of luminosity-weighted ages for the nuclear apertures\footnote{The exceptions are: 3C48, for which  we have included the information about the YSP ages from the extended aperture modelled by Canalizo and Stockton (2000); Centaurus A for which we have taken the YSP age information for the blue tidal stream extending up to 8~kpc from the nucleus;
and Fornax A for which we have used the age estimates for the blue globular clusters in the halo of the galaxy.} of the starburst radio galaxies. For comparison we also show the YSP age distribution for a complete sample of ULIRGs with redshifts $z < 0.13$ -- representing the extreme starburst population in the local Universe -- which
have been modelled using identical techniques by \citet{zaurin09}. Note that, in some apertures of some starburst radio galaxies it proved possible to model the optical spectra with YSP covering a wide range of ages. In such cases we have used the mean age over the range of models that provided good fits. Since the upper limiting ages in the latter group were often large ($>$1Gyr), this will tend to skew the  age distribution to larger ages.

Several features are apparent from Table 2 and Figure 1.

\begin{itemize} 
\item{\bf Nuclear apertures.} Considering the nuclear apertures (Figure 1 top), 9 (43\%) of sources have relatively young YSP ages: $t_{ysp} < 0.1$ Gyr. This group is interesting because the typical timescale of the starburst in a major galaxy merger is $\sim$0.1~Gyr \citep[e.g.][]{zaurin10}, and powerful extragalactic radio sources are generally considered to have maximum lifetimes in the range 0.01 -- 0.1~Gyr. For such objects the results are consistent with the starburst and AGN/jet activity occurring quasi-simultaneously, with little evidence for a major time lag between the starburst and the AGN/jet activity. On the other hand, there is also a group of 8 objects (36\% of sample) with well-determined older YSP ages ($0.3 < t_{ysp} < 3$~Gyr), in which the AGN/jet activity appears to have been triggered or re-triggered a significant period {\it after} the major star formation episode in the nuclear regions (see also the discussion in section 5.1). The latter group includes 3C305 and 3C293, which were discussed in detail in \citet{tadhunter05}. 

\item{\bf Extended vs. nuclear apertures.} Considering only the apertures with well-determined YSP ages, there is little evidence for a gradient in the YSP ages from the nuclear to the extended regions; in most cases the nuclear and extended apertures have luminosity-weighted YSP ages that are consistent given the (sometimes large) uncertainties. The exception is 3C459 for which the intermediate-aged YSP component becomes more significant off-nucleus \citep{wills08}, in common with the trend found for other ULIRGs in the local Universe \citep{zaurin09}.

\item{\bf Comparison with ULIRG YSP ages.} The overwhelming majority of nuclear apertures in the ULIRGs modelled by 
\citet{zaurin09} yield relatively young luminosity-weighted YSP ages: 88\%  have YSP ages $t_{ysp} < 0.1$ (Figure 1 bottom). This is consistent with the group of 9 radio galaxies indentified above with younger YSP ages. Indeed, four of the radio galaxies with younger YSP ages --- 3C48, PKS1345+12, PKS1549-79, 3C459 --- qualify as ULIRGs based on their far-infrared luminosities. Moreover, our modelling is consistent with young YSP ages for the one  other radio galaxy in our sample with a ULIRG-like far-IR luminosity --- PKS2135-20 --- even if the YSP age in that case cannot be determined with any accuracy, due to the presence of  strong nuclear AGN emission.

\item{\bf Links with level of AGN activity.} It is notable that {\it all} of the objects that have both highly powerful radio emission with $P_{5GHz} \ge 10^{26}$ W Hz$^{-1}$ and luminous [OIII] line emission with $L_{[OIII]} \ge 10^{35}$~W\footnote{This level of [OIII] luminosity corresponds to luminous quasar-like nuclear activity \citep[e.g.][]{zakamska03}.} 
have spectra consistent with young YSP ages ($t_{ysp} < 0.1$~Gyr). All 7 of these objects (33\% of the
sample) with powerful jets and luminous AGN are also luminous at far-IR wavelengths ($L_{ir} > 10^{11}$ W).
On the other hand, all but one of the objects (NGC612 is the exception) that have low radio powers with
$P_{5GHz}  < 10^{25}$~W Hz$^{-1}$ have spectra consistent with older, intermediate-age YSP. Thus, there appears to be a link between the level of AGN/jet activity and the age of the YSP, although the relationship is not perfect. Objects that do not fit in with the general trend include 3C321, which has highly luminous [OIII] and far-IR emission but intermediate age YSP, and 3C218 which has weak emission lines but much younger YSP. Interestingly both of the latter objects fall off the correlation between radio power and emission line luminosity: 3C321 is unusually emission line luminous for its radio power, whereas 3C218 has an unusually low emission line luminosity for its relatively high radio power, perhaps as a consequence of the effect of its rich
cluster environment boosting its radio luminosity. We further note that a substantial subset of radio galaxies with quasar-like levels of nuclear activity show no evidence for sgnificant recent star formation activity; these objects are discusssed further in section 5.4.
\end{itemize}

\subsection{Stellar masses}
\begin{table*}
\begin{tabular}{llllllll} 
\hline \\
Object &YSP Model &Mag &Ref &$M_{ysp}$ &$M_{osp}$ &$M_{tot}$ &YSP mass contr. \\
&(Age,E(B-V)) &Type & &($M_{\odot}$) &($M_{\odot}$) &($M_{\odot}$) & \\
\hline \\
PKS0023-26 &(0.03,0.9) &$r$' &1 &$6.0\times10^{10}$ &$2.9\times10^{11}$ &$3.5\times10^{11}$ &17\% \\
NGC612 &(0.05,1.2) &V$_T$ &2 &$1.6\times10^{11}$ &$7.5\times10^{11}$ &$9.1\times10^{11}$ &18\% \\
3C48 &(0.1,0)$^{*}$ &F555W &3 &$1.5\times10^{10}$ &$6.6\times10^{11}$ &$6.8\times10^{11}$ &2\% \\
Fornax A &-- &R$_J$ &4 &-- &$8.3\times10^{11}$ &$8.3\times10^{11}$ &-- \\
PKS0409-75 &(0.02,0.7) &$i$' &1 &$3.8\times10^{10}$ &$7.1\times10^{11}$ &$7.5\times10^{11}$ &5\% \\
PKS0620-52 &(0.7,0.0) &R$_c$ &7 &$2.3\times10^{10}$ &$1.6\times10^{12}$ &$1.6\times10^{12}$ &1\% \\
B20648+27 &(0.3,0.3) &m$_v$ &5 &$1.1\times10^{11}$ &$2.8\times10^{11}$ &$3.9\times10^{11}$ &28\% \\
3C213.1 &(0.6,0.0) &$r$' &6 &$2.1\times10^{10}$ &$1.7\times10^{11}$ &$1.9\times10^{11}$ &11\% \\
3C218 &(0.05,0.4) &R$_c$ &7 &$3.0\times10^{10}$ &$7.4\times10^{11}$ &$7.7\times10^{11}$ &4\% \\
3C236 &(0.05,1.4) &$r$' &6 &$2.6\times10^{11}$  &$6.6\times10^{11}$ &$9.2\times10^{11}$ &28\% \\
3C285 &(0.2,0.2) &$r$' &6 &$7.2\times10^{10}$ &$3.5\times10^{11}$ &$4.2\times10^{11}$ &17\% \\
Centaurus A &-- &R$_T$ &8 &-- &$1.4\times10^{11}$ &$1.4\times10^{11}$ &-- \\
PKS1345+12 &(0.05,0.8) &R &11 &$7.5\times10^{10}$ &$8.3\times10^{11}$ &$9.0\times10^{11}$ &8\% \\
3C293 &(2.0,0.4) &$r$' &6 &$1.6\times10^{11}$ &$1.2\times10^{11}$ &$2.8\times10^{11}$ &57\% \\
3C305 &(0.4,0.6) &$r$' &6 &$1.1\times10^{11}$ &$4.3\times10^{11}$ &$5.7\times10^{11}$ &19\% \\
3C321 &(0.6,0.2) &$r$' &6 &$3.0\times10^{10}$ &$5.1\times10^{11}$ &$5.4\times10^{11}$ &6\% \\
PKS1549-79 &(0.05,0.4) &R &9 &$5.0\times10^9$ &$1.6\times10^{11}$ &$1.6\times10^{11}$ &3\% \\
PKS1932-46 &-- &$r$' &1 &-- &$3.3\times10^{11}$ &$3.3\times10^{11}$ &--- \\
3C433 &(0.05,0.7) &V$_{25}$ &10 &$3.2\times10^{10}$ &$4.3\times10^{11}$ &$4.7\times10^{11}$ &7\% \\
PKS2135-209 &(0.2,0.2) &$i$' &1 &$5.6\times10^{10}$ &$3.1\times10^{11}$ &$3.7\times10^{11}$ &15\% \\
3C459 &(0.05,0.2) &$r$' &1 &$1.3\times10^{10}$ &$1.6\times10^{11}$ &$1.7\times10^{11}$ &8\% \\
\hline \\
\end{tabular}
\caption{Estimated stellar masses for starburst radio galaxies. The second column gives the
age and reddening of the YSP in the model used to derive the masses; the third column gives the
type of magnitude used to scale the models, with reference to the magnitude in column four; the
fifth, sixth and seventh columns give the YSP, OSP and total stellar masses (12.5~Gyr OSP age
assumed for all cases, except the higher redshift objects PKS0409-75 and PKS2135-20, for which a 7.0 Gyr
OSP was assumed); and the final column
gives the proportional contribution of the YSP to the total stellar mass.  Note that in the
cases of Fornax A, Centaurus A and PKS1932-46 we do not have an estimate of the
proportional contribution of the YSP to the total stellar light. In these cases we have assumed 
that the starlight is dominated by old stellar populations. $^{*}$ For 3C48 we have assumed that model B8 from 
\citet{canalizo00} is representative of the stellar populations in the host galaxy as a whole.   
Key for photometry references:
1. Ramos Almeida et al. (2010); 2. RC2; 3. Boyce et al. (1999); 4. Persson et al. (1979);
5. Gonzalez Serrano \& Caballo (2000); 6. Sloan Digital Sky Survey DR6 (magnitude
derived from CMODEL fit); 7. Govoni et al. (2000); 8. Lauberts et al. (1989);
9. Drake et al. (4004); 10. Smith \& Heckman (1989); 11. Kim et al. (2002).
 } 
\end{table*}

\begin{figure} 
\psfig{file=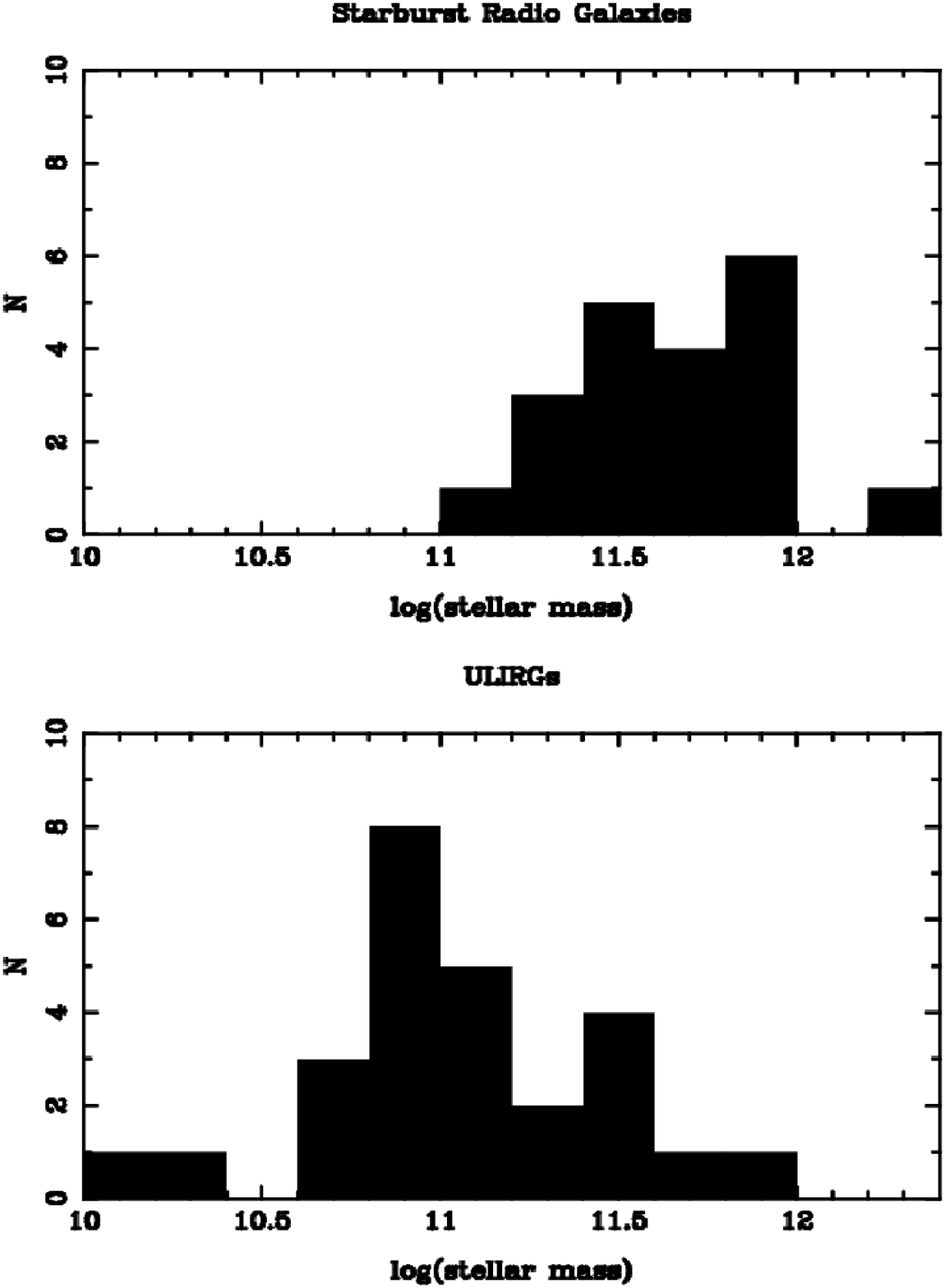, width=8cm}
\caption{The total stellar masses of starburst radio galaxies (top), compared with
those of ULIRGs with $z < 0.13$ from the complete sample (CS) 
of \citet{zaurin09} (bottom). See the text for details of how the masses have been calculated.}
\end{figure}

In order to compare the starburst radio galaxies with other types of star forming galaxies in the local
Universe it is important to consider their stellar masses. We have estimated the total stellar masses  by using the monochromatic continuum fluxes derived from the best available total photometric magnitudes to scale the stellar masses derived from our spectral synthesis modelling of the spectra. In each case we used the best-fitting YSP+OSP model to derive the stellar masses for the nuclear spectroscopic aperture, then assumed that the same model was appropriate for the entire stellar halo of the galaxy sampled by the photometric aperture. This assumption is justified on the basis that our spectra show little evidence for major gradients in the stellar populations across the halos of most of the galaxies in our sample 
\citep[e.g.][]{holt07}. Although the major source of uncertainty in this procedure is likely to be related to the choice of a single stellar model for each galaxy, experiments involving the adoption of extreme models (e.g. assuming that the halos are dominated by OSPs) suggest that the derived total stellar masses are accurate to within a factor of two. 

The results for the starburst radio galaxies are presented in Table 4 and Figure 2. They demonstrate that most starburst radio galaxies are giant elliptical galaxies in terms of their total stellar masses. Considering the galaxy mass function derived by \citet{cole01}\footnote{The mass function given in \citet{cole01} has a characteristic
mass: $m_{*}=1.4\times10^{11}$~M$_{\odot}$, adapted to our cosmology}, the majority of starburst radio galaxies are super-m$_{*}$: the mean and median stellar masses are $5.6\times10^{11}$~M$_{\odot}$ (4$\times$m$_{*}$) and $4.7\times10^{11}$~M$_{\odot}$ (3.3$\times$m$_{*}$). These results are consistent with those derived for the general population of powerful radio galaxies at all redshifts \citep[e.g.][]{dunlop03,volmerange04,seymour07,inskip10}. For example, based on
the near-IR photometry measurements presented in \citet{inskip10}, the general population of nearby ($z < 0.5$) 2Jy radio galaxies has stellar masses in the range
$2\times10^{11} < M_{stellar} < 2\times10^{12}$~M$_{\odot}$. However, as highlighted in \citet{wills08}, it is dangerous to generalise, and a minority of radio galaxies have stellar masses that are significantly lower ($\sim m_{*}$).

Figure 2 also shows a comparison between the total stellar masses of starburst radio galaxies (top panel) and nearby ULIRGs from the $z < 0.13$ complete sample of \citet{zaurin09} (bottom panel), derived using identical techniques. In making this comparison it is important to recognise that the estimated total stellar masses of the ULIRGs have been derived using OSP+YSP models \citep[modelling combination I in][]{zaurin09}, which tend to maximise the OSP combination (and hence the total stellar masses), whereas in fact the nuclear spectra of many of ULIRGs can be adequately modelled by combinations of intermediate age and young YSP, without any contribution from an OSP \citep[combination III in][]{zaurin09}, similar to the case of 3C459 amongst the starburst radio galaxies \citep{wills08}. Despite the fact that the masses of ULIRGs in Figure 12 may be over-estimated, most radio galaxies have total stellar masses that are greater than those of most ULIRGs, although there is a significant overlap between the two distributions. Using a K-S two sample test we can reject the null hypothesis that
the two samples have the same mass distributions at the  $>$99.5\% level
of significance. Overall, this comparison provides a clear demonstration of the fact that, although some radio galaxies {\it are} ULIRGs and others may have {\it evolved from} ULIRGs \citep[see][]{tadhunter05}, not all ULIRGs can evolve
into radio galaxies and vica versa; only the most massive 50\% of ULIRG systems are capable of harbouring powerful radio jets. In this context it is notable that, out of the complete sample of 26 $z < 0.13$ ULIRGs investigated by \citep{zaurin09}, only one is a radio galaxy, and that object --- PKS1345+12 (also in our sample of starburst radio galaxies) --- has the largest total stellar mass of all the ULIRGs.

\section{Discussion}

\subsection{Triggering radio galaxy activity in galaxy mergers}

It is clear that many of the optical morphological features of starburst radio galaxies are consistent with the idea that both the AGN/jet and starburst activity have been triggered as the result of
the gas infall associated with galaxy mergers. Hydrodynamical simulations of mergers reveal that they are complex events, and it can take a gigayear or more for
the merger remnant to relax and achieve the appearance of a normal galaxy. The main gas infall and star formation 
are predicted to occur in two major peaks: one at or just after the first peri-centre passage of the merging nuclei; the other as the two nuclei finally coalesce \citep[e.g.][]{mihos96}. The relative intensity and separation in time of these two peaks depends on several factors such as the morphologies, mass ratios and 
gas contents of the precursor galaxies, the merger geometries, and the importance of feedback processes associated with the AGN and starbursts \citep{mihos96,dimatteo05,dimatteo07,cox08,johansson09}. 

Given that most starburst radio galaxies have massive old stellar populations in addition to their YSP, it is likely their precursors had signifcant stellar bulges. Moreover, the
relatively large masses and proportional mass contributions of the YSPs (see Table 5)
suggests that at least one of the precursors must have been gas-rich. In such circumstances the models predict that the second star formation peak --- associated with the coalescence of the two nuclei --- is likely to be the most intense \citep{mihos96}. If the AGN feedback effects are important, and the AGN is triggered concurrently with the starburst, then the star formation will
be halted shortly after ($<$0.1~Gyr) the final coalescence of the nuclei by powerful AGN-induced winds.
The time lag between the two star formation peaks is of order 0.3 -- 1.5~Gyr, while the major phase of activity associated with the second (more intense) peak  is expected to last of order 0.1~Gyr. However, star formation and AGN activity may not only be associated with the two main peaks of activity, but may also occur --- albeit at reduced intensity --- at other stages, for example, as the tidal debris rain back down on the merger
remnant following the coalescence of the two nuclei. Many of the most recent simulations demonstrate
that the 
fuel supply to the central supermasive black holes can be highly irregular over the course of a merger \citep{dimatteo07,cox08,johansson09}. 

It is important to emphasise that, as well as the delivery of gas into the {\it general vicinity} of the supermassive back hole (within $\sim$100~pc -- the typical resolution of the hydrodynamical sigmulations), there may be other timescales involved in the triggering of the activity. These include the time required for the supermassive black holes from the precursor galaxies to merge, and the time required for the accreted gas to lose sufficient angular momentum to move close enough to the remnant supermassive black hole to fuel the activity \citep[see discussion in][]{tadhunter05}. It has also been suggested that the powerful winds
associated with the circum-nuclear star formation activity may disrupt the flow of gas into the nuclear regions and delay the onset of AGN activity \citep{davies07,wild10}.  Unfortunately the current
generation of hydrodynamical simulations do not have sufficient resolution to include these processes in any detail. Therefore
in what follows we will assume that the black hole accretion and star formation rates are as (crudely) predicted by the simulations, with the caveat that the reality is likely be more complex. 

\begin{figure*} 
\psfig{file=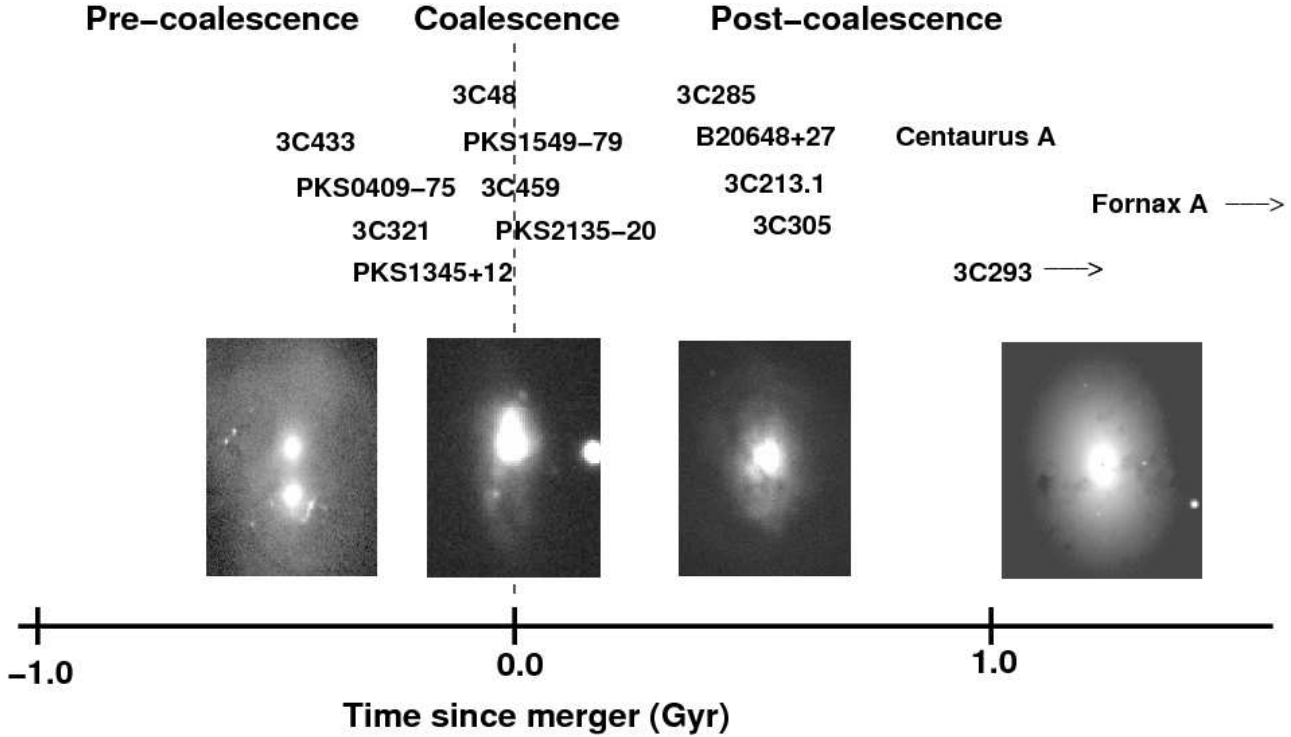, width=17cm}
\caption{An evolutionary sequence for starburst radio galaxies triggered in galaxy mergers.}
\end{figure*}

We now consider whether, based on their YSP properties and optical morphologies, the objects discussed in this paper are consistent with the  merger scenario for the triggering of the activity. Most of the starburst radio galaxies can be fitted into one of the following three categories, each corresponding to a particular phase of a gas-rich merger\footnote{N.B. Only objects with good YSP age estimates, that fit clearly in the merger sequence in terms of their ages and morphologies, are included as examples. Misfits are discussed later in this section.}. This merger sequence is also illustrated in Figure 3. 

\begin{itemize}
\item {\bf Pre-coalescence.} These double nucleus systems are observed after the first peri-centre passage, but immediately before (within $\sim$0.1~Gyr) the coalescence of the two nuclei, as they move together. All of these objects have large AGN/jet and far-IR luminosities, reflecting the level of AGN and starburst activity associated with the gas infall expected at this stage of a merger. The YSP in these systems represent a combination of the younger YSP formed as a consequence of the fresh gas infall associated 
with coalescence, and intermediate age stellar populations formed in the first peak of star formation, around the time of the first peri-centre passage. The relative importance of the latter two stellar populations will depend on the details of the encounter,
and the extent to which the most recent episode of star formation is obscured by dust. Alternatively, any intermediate age stellar populations could be associated with the captured disks of the precursor galaxies. Examples: PKS0409-75, PKS1345+12, 3C321 and 3C433.

\item {\bf Coalescence.} These single nucleus systems are observed in the 0.1~Gyr period immediately following the coalescence of the two nuclei. For major mergers involving gas-rich galaxies with significant bulges, the models predict that this should represent the most active phase of starburst and AGN activity. Due to the fact that the most recent phase of star formation activity has been particularly intense, and is likely to
dominate (in flux) over any previous generations of stars, the luminosity-weighted ages in the nuclear regions of these objects are relatively young ($<$0.1~Gyr). The far-IR, [OIII] and radio luminosities are large in this phase, reflecting the high level of AGN and starburst activity associated with the large concentrations of gas and dust in the nuclear regions; in some cases (depending on the
details of the merger) the source may attain a ULIRG-like luminosity. Examples: 3C48\footnote{Note that, although it has been proposed that this source contains a secondary nucleus --- visible in HST images, the putative secondary nucleus is much closer to the primary nucleus (only ~1kpc in projection) than in the cases of the pre-coalescence sources listed above. Therefore we list it as a coalescence case.}, PKS1549-79, PKS2135-209 and 3C459. 

\item {\bf Post-coalescence.} These single nucleus systems still bear the morphological hallmarks of a merger in the recent past (tidal tails, dust lanes, distorted outer isophotes etc.), but they have intermediate age YSP ($0.2 < t_{ysp} < 2.5$~Gyr), consistent with them being observed a significant period after the main merger-induced starburst (i.e. the YSP represent post-starburst stellar populations). This is the scenario outlined in \citet{tadhunter05} and \citet{emonts06}. Unless the AGN/jet activity lasts considerably longer than the 0.1~Gyr maximum lifetime that is typically estimated for powerful extragalactic radio sources \citep[e.g.][]{leahy89}, there must have been a time lag between the main merger-induced starburst and the triggering of the current phase of activity. Examples: B20648+27, 3C213.1, 
3C285, 3C293, 3C305, Centaurus A, Fornax A.
\end{itemize}

Just as it appears likely that radio-loud AGN activity can be triggered at different stages during a merger, it is also possible that there is more than one episode of AGN/jet activity in the course of a merger. For example, an individual system
may undergo a particularly powerful phase of AGN/jet activity as the nuclei coalesce, close to the peak of starburst activity (the coalescence phase above), but the AGN/jet activity may also
be triggered (or re-triggered) earlier or later in the merger sequence, depending on the details of the radial gas flows in the merger. Indeed, as noted above, six of the starburst radio galaxies in our sample 
show evidence for re-triggered radio source activity in the form of high surface brightness inner radio structures
and more diffuse and extended outer radio structures (3C213.1, 3C218, Cen A, 3C236, 3C293, PKS1345+12). Moreover, in the particular case of 3C236, \citet{odea01} have argued for multiple phases of jet activity, based on
both the double-double morphology of its radio source, and the evidence for two major epochs of star formation
in its host galaxy.

We emphasise that not all of the starburst radio galaxies can be readily accommodated
within the merger scheme outlined above; there are some prominent misfits. Most notably, the central
cluster galaxies 3C218 and PKS0620-52 do not show clear morphological signs of major mergers, and 3C218 is also unusual in having relatively young YSP ($t_{ysp}\sim$0.05~Gyr) but low emission line and far-IR luminosities (the age of the YSP in PKS0620-52 is not well-determined). PKS0023-26 may also represent an ambiguous case in the sense that it lies at the heart of a rich cluster of galaxies, and has a peculiar amorphous outer envelope that is difficult to classify in terms of the merger sequence (although a merger cannot be entirely ruled out). 

In the case of the one system in our sample that shows evidence for large-scale, and relatively settled, gaseous disk --- NGC612 --- it is not clear whether the AGN/jet activity has been triggered by interactions with massive galaxies on a large scale in the wider galaxy group (as may be evidenced by the HI observations), or by recent mergers/interactions with closer companion galaxies \citep[see][]{emonts08a}. The latter seems more likely given that NGC612 shows an optical shell structure, along with YSP that are
much younger ($t_{ysp} < 0.1$~Gyr) than the estimated time since its closest approach to the 
the massive companion galaxy NGC619 ($\ge 1$~Gyr). Clearly, the 
possibility of multiple interactions and mergers in galaxy groups can complicate the interpretation of the galaxy morphologies and
YSP ages in terms of a simple merger sequence.

Finally we note that, while the properties of the most luminous starburst radio galaxies -- in particular, the ULIRG-like coalescence systems --  are consistent with triggering in major (similar mass) gas-rich mergers, it is difficult to rule out minor mergers as the trigger for some of the other systems. Indeed, it has been argued that the large-scale features of Centaurus A are consistent with a minor (1:10) merger between the host radio galaxy and a smaller disk galaxy
\citep{malin83a}.

\subsection{Do cooling flows play a significant role?}

As noted in the previous section, some of the starburst radio galaxies that are situated close to the centres
of rich clusters of galaxies are not readily accommodated in the merger sequence. For such objects
cooling flows\footnote{By cooling flows we mean here the cooling of the large-scale hot ($\sim10^7$ -- $10^8$~K)
X-ray emitting gas of the cluster to cooler phases of the ISM in the central cluster galaxy, where
it can form stars and/or be accreted by the central black holes. The direct accretion of the hot gas itself
by the central black hole will be discussed in the following section.} associated with the hot X-ray emitting gas may provide an alternative triggering mechanism for both the visible star formation and the AGN/jet activity. 
This mechanism is supported by the  irregular morphologies and kinematics of the emission line gas in central cluster galaxies hosting radio sources \citep{tadhunter89,baum92}. Moreover, while the high velocity dispersions of the galaxies in the centres of massive
galaxy clusters can hamper gas accretion via major, gas-rich mergers, there is no such problem for the accretion of
warm/hot gas via cooling flows, since the cooling gas will naturally fall towards the centre of the the cluster potential
well.

Although star formation has been discussed as a potential
sink for the cooling gas, until recently the apparently large differences between the estimated hot gas cooling rates and the star formation rates in the central cluster galaxies suggested that only a small fraction of the the cooling gas ends up forming stars \citep[e.g.][]{mcnamara89}. However, spectroscopic observations with the new generation of X-ray satellites have led to a major downward revision in estimates
of the hot gas cooling rates, so that they are now much closer to the star formation rates. Therefore, it is plausible that a significant fraction of the cooling gas does in fact form stars  \citep[see][]{rafferty06}. In the cases of the three cooling flow candidates in our sample, based on their infrared luminosities and the relation of \citet{kennicutt98}, the star formation rates
are 5, 4 and 25~M$_{\odot}$ yr$^{-1}$ for PKS0620-52, 3C218(Hydra A) and PKS0023-36 respectively. Of these three, only 3C218 has 
published high quality X-ray observations, and its hot gas cooling rate of $16\pm4$~M$_{\odot}$ yr$^{-1}$ \citep{rafferty06} proves to be within a factor of 4 of the star formation rate. 

Given the agreement between its star formation and hot gas cooling rates, 3C218 is one of the best candidates for an
object in which the activity has been triggered by the warm/cool gas condensing out of a cooling flow.   However, even in the case of 3C218 we cannot entirely rule out the idea that the activity has been triggered in a galaxy merger or interaction, since it is possible for central cluster galaxies to undergo mergers that could, potentially, form  star forming gaseous disks similar to that associated with the dust lane in the central regions of 3C218 \citep{ramos10}. Although such disks (and the merging galaxies that produced them) may represent only a small fraction of the total masses of the cD galaxies, the associated gas infall rates may be sufficient to fuel the AGN/jet activity in the nuclei on the requisite timescales. Note that the large-scale morphological signatures of galaxy mergers (e.g. tidal tails, fans, shells) are more difficult to detect against the light of the massive stellar haloes of the
central cluster galaxies than they are in lower mass elliptical galaxies; the tidal features are also likely to be erased on a relatively
short timescale by ongoing tidal interactions between all the galaxies in the dense central regions of the galaxy clusters.


\subsection{Triggering the activity in weak line radio galaxies}

As a class, low-power FRI radio galaxies are almost invariably classified as weak line radio galaxies (WLRG) at optical wavelengths. However, a  significant proportion of FRII galaxies also have WLRG spectra, with 
[OIII]$\lambda$5007 and mid-IR luminosities that are an order of magnitude lower than narrow- and broad-line radio galaxies (NLRG/BLRG) of comparable radio power 
\citep{tadhunter98,dicken09}. On the basis of their nuclear X-ray and emission line properties, and the way that their emission line luminosities and absolute magnitudes correlate with radio power, it has been suggested that the nuclear accretion rate or accretion mode in WLRG is distinct from that in their strong-lined radio galaxy counterparts \citep[e.g.][]{hardcastle07,buttiglione10}. In particular, it has been argued that WLRG represent objects in which the AGN/jet activity is fuelled via Bondi accretion of the hot X-ray-emitting ISM, while strong-lined radio galaxies represent objects in which the activity is fuelled by the accretion cooler phases of the ISM. For low power radio galaxies the Bondi accretion hypothesis is supported by correlations between galaxy/black hole mass and the fraction of galaxies that are radio-loud, the relatively large proportion of galaxies that are radio-loud at large galaxy/black hole mass (suggesting a high duty cycle), and the fact that the energy input of radio jets fuelled by Bondi accretion is sufficient to balance radiative losses of the hot X-ray emitting haloes of the host giant elliptical galaxies
\citep{best05,allen06,best06}.

Significant star formation is not expected in the case of triggering via Bondi accretion of the hot ISM. Therefore it is interesting that a significant subset of the starburst radio galaxies 
are classed WLRG (NGC612, PKS0620-52,  3C213.1, 3C218, 3C236, 3C293, Fornax A, Centaurus A). Although some
such objects (PKS0620-52, 3C218) are candidates for fuelling via cooling flows (see section 5.2 above),
more than 50\% of the WLRG in our sample of starburst radio galaxies can be classified as post-coalescence or late post-coalescence systems, with evidence for a rich ISM in the circum-nuclear regions
in the form of circum-nuclear dust lanes, as well as other signs that they have been involved in mergers in the recent past. This suggests a possible alternative to fuelling via Bondi accretion of the hot ISM: accretion of the cool ISM at a slow rate from the debris disks of galaxy mergers. As the disks settle it is expected
that the shocks associated with the gaseous dissipation process will lead to a net infall of gas, and also produce a LINER-type emission line spectrum, similar to those observed in WLRG  \citep{dopita97}.

In the later post-merger stages the debris disks may also act as large reservoirs of {\it potential} fuel for the AGN/jet activity, but external stimuli, such as interactions with neighbouring galaxies or the accretion of 
satellite galaxies, may be required to perturb the disks and induce radial gas infalls that are sufficient
to trigger the AGN. This might help to explain late post-coalescence systems such as the iconic FRI radio galaxies Centaurus A and Fornax A, which appear to have been triggered a substantial period after ($\sim$2 -- 3~Gyr after in the case of Fornax A) the mergers associated with the formation of the dust lanes. In such
cases, the original merger may not have triggered the current phase of nuclear activity directly, but
rather delivered a reservoir of cool gas that was then perturbed by more minor galaxy interactions to produce
the observed level of activity. In this context it is notable that Centaurus A shows evidence for an ongoing
satellite galaxy accretion event in the form of a blue tidal stream \citep{peng02}.

Overall, the assumption that {\it all} WLRG are fuelled by Bondi accretion of the hot ISM is likely to represent an over-simplification; for a significant fraction -- including some FRI radio galaxies -- slow accretion of cooler gas from the debris disks of galaxy mergers represents a viable alternative mechanism. This is consistent with the X-ray and mid-IR evidence for compact, cool dust/gas structures in the nuclear regions of a minority of WLRG, including Centaurus A \citep{evans04,ramos11}.

\subsection{Non-starburst radio galaxies}

The discussion above has concentrated on starburst radio galaxies that show strong evidence for recent star formation activity at optical wavelengths. However, such objects comprise only a small proportion ($\sim$15 -- 25\%) of the full population of powerful radio-loud AGN in the local Universe. Clearly it is important to consider how the AGN/jet activity is triggered in the 75 -- 85\% majority of powerful radio galaxies that do not show strong signs of optical star formation activity. Note that many of the latter belong to the class of strong-lined FRII radio sources in which the AGN/jet activity is generally considered to be triggered by accretion of the cooler phases of the ISM \citep{hardcastle07,buttiglione10}.

One possibility is that the non-starburst radio galaxies are objects in which the AGN/jet activity has been triggered or re-triggered in the late post-coalescence phase, $>$1~Gyr after the starburst associated with the coalescence of the nuclei in the merger. As argued in \citet{wills02}, after $\sim$1~Gyr it can be difficult to detect a merger-induced starburst at optical wavelengths against the light of the giant elliptical galaxy host, especially if the YSP comprises a relatively small fraction of the total stellar mass and/or suffers a moderate amount of extinction. In the late post-coalescence phase, the morphological signs of mergers also become more difficult to detect, especially for radio galaxies at intermediate/high redshifts. The main problem with this explanation is that, for strong-lined, non-starburst FRII radio galaxies, the moderate rates of gas infall associated with the debris disks may not be sufficient to fuel the prodigious nuclear activity. Indeed, 
many of the starburst radio galaxies that we identified above as post-coalescence systems are WLRG. 

Alternatively, it is possible that the activity in the non-starburst radio galaxies has been triggered by close encounters with gas-rich companion galaxies, or following the first peri-center passages of the merging nuclei in galaxy mergers. In the latter case, hydrodynamical simulations show that the tidal forces following the first encounter induce significant radial infalls of gas into the nuclei for
an extended period around (\citep[$\sim$0.5 -- 1.0~Gyr:][]{springel05,johansson09}) the time of closest approach. Such infalls may capable of fuelling the prodigious AGN/jet activity in strong-lined radio galaxies. However, in cases in which the merging nuclei
have significant bulges (highly likely in the case of radio galaxies), only a relatively low level of star formation is expected at this stage. This mechanism may be supported by the finding of a relatively high incidence of tidal bridge features and tidally distorted companion galaxies in deep imaging observations of non-starburst radio galaxies in the 2Jy sample \citep{ramos10}; it is further supported by detailed studies of individual radio galaxies in interacting groups \citep{inskip07,inskip08}.

A final possibility is that non-starburst radio galaxies are triggered in relatively minor mergers (1:3 or less), or in major mergers that are relatively gas poor (``dry''). In such cases relatively low levels of star formation are expected, which are likely to be difficult to detect against the light of the old stellar populations in the host galaxies. However, without further theoretical work, it is not clear whether minor or gas-poor mergers would be capable of delivering sufficient gas to the nuclear regions to fuel the quasar-like levels of nuclear activity detected in some non-starburst radio galaxies.

\subsection{The radio structures of starburst radio galaxies}

An interesting aspect of starburst radio galaxies is that they show a high incidence of unusual radio structures: our sample includes several compact CSS/GPS sources, as well as sources with relatively bright steep spectrum core structures, diffuse outer haloes and double-double structures.
What (if any) is the relationship between these unusual radio structures and the presence of young stellar populations in the host galaxies? 

As already discussed above, double-double sources, and sources with compact high surface brightness inner structures combined with diffuse outer haloes, may represent cases in which the radio jet activity has been re-triggered (but see Morganti et al. 1999 and 
Wise et al. 2007 for counter-arguments in the cases of Centaurus A and Hydra A). Given  the complexity 
of the gas infall histories of major gas rich mergers, it is certainly plausible that each merging system undergoes more than one phase of AGN/jet activity, thus explaining the presence of such sources in our sample.

Considering the compact (CSS/GPS) radio sources, there is direct observational evidence from measurements of hotspot advance speeds that CSS/GPS radio sources are relatively youthful ($10^4$ -- $10^6$~yr). However, since the CSS/GPS radio sources are generally estimated to be much younger than the YSPs detected in their
host galaxies ($t_{ysp} \sim 10^7$ -- $10^9$~yr), the youth of the compact sources does not necessarily help to explain their relatively high rate of occurrence in our sample of starburst radio galaxies. 

Alternatively, the high incidence of CSS/GPS sources in the starburst radio galaxies sample may be the consequence of an observational selection effect as follows. We expect a relatively rich and dense ISM to be present in the nuclear regions of merging systems, especially around the
time of nuclear coalescence. A radio source triggered in a merger will interact particularly strongly with this rich ISM in the early stages of the radio source history, as the
jets expand through the central regions of the host galaxies; direct evidence for strong jet-cloud interactions in young radio sources is provided by their extreme emission line kinematics \citep{holt08}. The strong interactions between the jets and the rich ISM associated with the mergers will not only result in extreme emission line kinematics, but may also affect the conversion of jet power into radio luminosity, boosting the radio luminosities of sources. There is already evidence that interaction with the relatively dense, hot X-ray haloes associated with clusters of galaxies boosts the radio luminosities
of jets for a given jet power \citep{barthel96}, and it is plausible that there will be a similar boosting effect when the jets strongly interact with the (cooler) ISM in the central regions of merger remnants. Indeed, strong enhancements in the radio emission are observed at the sites of interactions between radio jets and warm emission line clouds in the haloes of radio galaxies in the local Universe \citep[e.g.][]{vanbreugel85,vanbreugel86,fosbury98,tadhunter00}. For a given intrinsic jet power, this flux boosting will lead to the compact radio sources that are
triggered in young, star forming merger remnants being preferentially selected in flux-limited radio surveys. This
in turn could explain the relatively high rate of occurrence of compact radio sources amongst the starburst radio galaxies, as well as the bias of the compact sources towards relatively young YSP ages ($t_{ysp} < 0.1$~Gyr; see section 4.1).

The interaction of the jets with the richer gaseous environments present in merger remnants could also help to explain the relatively high incidence of extended radio sources with compact steep spectrum cores in our sample, since such interactions have the potential to boost the radio emission from
the jets, even if the radio lobes are well ouside the central regions of the galaxies.

\section{Conclusions and future work}

In this paper we have discussed the properties of the young stellar populations (YSP) in the 
$\sim$15 -- 25\% of powerful radio
galaxies that show strong evidence for recent star formation activity at optical wavelengths. Combined with information about the
morphologies of the host galaxies, the YSP properties of most of these starburst radio galaxies are consistent with
the triggering of both starburst and AGN/jet activity in galaxy mergers in which at least one of the merging galaxies
is gas-rich. However, the triggering of the AGN/jet activity is not confined to a single evolutionary phase of galaxy mergers.
While in a significant subset of objects the AGN/jet activity has been triggered within 0.1~Gyr of the coalescence of merging nuclei, 
close to the expected peaks of the merger-induced starbursts, many objects are observed in the post-coalescence phase, $>$0.2~Gyr after
the starburst peaks. In the former group the triggering of the activity can be readily identified with the major infalls of gas that
are predicted to occur around the time of coalescence. On the other hand, in the latter group the nature of the link between the 
triggering of the AGN/jet and the original starburst-inducing mergers is less clear: late-time infall of merger debris, settling of
the debris disks to an equilibrium configuration, and perturbation of the debris disks by minor mergers and encounters, are
all possibilities for the triggering of the activity in these, generally lower luminosity, objects.

Our results clearly demonstrate that luminous radio-loud AGN {\it can} be triggered close to the peaks of major gas-rich mergers. However, the relatively low incidence of radio-loud AGN in nearby ULIRGs suggests that, even under similar conditions of gas infall, it is not inevitable that radio-loud
AGN activity {\it will} be triggered in such mergers. Indeed our comparison of the stellar masses of ULIRGs and starburst radio galaxies suggests that
the triggering of powerful radio-loud AGN activity is most likely to occur in the most massive merger remnants, consistent with other studies of AGN host galaxies that provide evidence for a link between host galaxy mass (and hence black hole mass) and the incidence of the radio-loud AGN population \citep[e.g.][]{lacy01}.

Overall, if we also take into consideration the evidence that some radio-loud AGN are triggered by 
the accretion of gas from the hot X-ray
haloes, either via cooling flows or the direct (Bondi) accretion of hot gas, it appears unlikely that radio-loudness is solely a consequence
of a  particular mechanism for the delivering of the fuel supply to the nuclear regions. Rather, the ability to produce powerful relativistic
jets is more likely to be related to the intrinsic properties of energy generating regions (e.g. the spin and/or mass of the supermassive
black hole).

Finally, we emphasise that a considerable fraction of powerful radio galaxies ($\sim$75 -- 85\%) do not show evidence for
energetically significant starburst activity at either optical or MFIR wavelengths. Therefore, in order to complete our understanding of
triggering of the activity in radio-loud AGN, in the future it will be important to compare the host galaxy properties and environments
of the non-starburst majority of radio galaxies, with those of the starburst minority discussed in this paper.

\section*{acknowledgments} CRA, JH and KJI acknowledge support from the UK Science and Technology Facilities Council
(STFC), De Nederlandse Organisatie voor Wetenschappelijk Onderzoek (NWO), and the Emmy Noether programme of
the German Science Foundation (DFG) respectively. We thank the anonymous referee for useful
comments that have helped to improve the manuscript. This research has made use of the NASA/IPAC Extragalactic Database (NED)
which is operated by the Jet Propulsion Laboratory. Based on observations made
with ESO Telescopes at the La Silla and Paranal observatories under programmes 70.B-0663(A), 71.B-0320(A), 078B-0660(A).

\newpage\noindent
{\large\bf Appendix I: descriptions of individual objects}
 
The appendix will be available in the published version of the paper.
 
\end{document}